\documentstyle[11pt]{article}
\newcommand{\tb}{\left(\frac{\tau}{\beta}\right)}
\newcommand{\A}{{\cal A}}
\newcommand{\M}{{\cal M}}
\newcommand{\la}{{\frac{\Lambda}{6}}}

\begin{document}
\normalsize

 {\LARGE{\bf
 \centerline {Interpretation of the Siklos solutions}
 \centerline { as exact gravitational waves}
 \centerline { in the anti-de Sitter universe}}}
 \vspace{25mm}
 \centerline {Ji\v r\' \i\  Podolsk\' y}
 \centerline {Department of Theoretical Physics,}
 \centerline { Faculty of Mathematics and Physics, Charles University,}
 \centerline {V Hole\v sovi\v ck\'ach 2, 180 00 Prague 8, Czech Republic }

 \centerline {E-mail address: podolsky@mbox.troja.mff.cuni.cz}

 \vspace{10mm}
 \centerline {December 10, 1997}

 \vspace{25mm}

\begin{abstract}
The Siklos class of solutions of Einstein's field
equations is investigated by analytical methods. By studying the
behaviour of free particles we reach the conclusion that the
space-times represent exact gravitational waves propagating in the
anti-de~Sitter universe. The presence of a negative cosmological constant
implies that the `background' space is not asymptotically flat
and requires a `rotating' reference frames in order to fully
simplify and view the behaviour of nearby test particles. The
Kaigorodov space-time, which is the simplest
representative of the Siklos class, is
analyzed in more detail. It is argued that it may serve as a
`cosmological' analogue  of the well-known homogeneous {\it pp}-waves
in the flat universe.

\vspace{3mm}
\end{abstract}

\newpage
\section {Introduction}

The first class of exact solutions representing gravitational waves
in general relativity was found by Brinkmann in 1923  \cite{Brink}. The
metrics were later discovered independently by several authors
(cf. \cite{KSMH}) including Robinson  who in 1956 recognized
their physical meaning --- these metrics are now known as {\it pp}
-waves. In 1925, Beck \cite{Beck} discovered  cylindrical gravitational
waves which were later studied by Einstein and Rosen \cite{ER}.
In the beginning of the 1960's, the introduction of new geometrical
concepts and methods  (algebraic classification,
gravitational ray optics, concept of the news function,
spin coefficients etc.)  had an important influence on
finding new exact radiative solutions. These solutions, such as the
plane-fronted waves \cite{BPR}-\cite{EK}
or the Robinson-Trautman `spherical' waves \cite{RT2}, are now
considered as standard `prototypes' of exact gravitational waves.

An important step in treatment of gravitational radiation
within the full non-linear general relativity was made by Penrose.
His concept of a smoothly
asymptotically flat space-time (cf. \cite{PR} and references therein)
represents a rigorous geometrical framework for the
discussion of gravitational radiation from  {\it spatially isolated}
sources. Moreover, the case of finite sources has an  astrophysical
relevance so that most of the work on gravitational radiation
has been concerned with space-times which are either
asymptotically flat (in some directions at least), cf. \cite{BS}, or contain
flat regions explicitly as, e.g., in the case of colliding plane gravitational
waves (see \cite{Grif} for a comprehensive review).

On the other hand, in the last two decades new exact solutions representing
`cosmological' gravitational waves in non-asymptotically flat models were
found and analyzed, for example in \cite{Gowdy}-\cite{AlGrif2}  and elsewhere
(cf. \cite{CCM, BGM, Verd} for a review of the main works).
Some of these solutions (usually admitting two spacelike Killing
vectors) can be interpreted
as spatially inhomogeneous models in which the homogeneity of
the universe is broken due to gravitational waves.
They may serve as exact models of the propagation of primordial
gravitational waves and may be relevant for the (hypothetical)
cosmological wave background.

In this work we concentrate on the physical interpretation of the
class of exact type N solutions with a negative
cosmological constant $\Lambda$ found by Siklos \cite{Siklos}.
In general, these solutions admit only one Killing vector.
Moreover, since $\Lambda<0$, we deal with gravitational waves `in'
an everywhere curved anti-de Sitter universe
(the de Sitter and anti-de Sitter universes are the simplest
natural cosmological `background' space-times for solutions with
$\Lambda$ because they are conformally flat and they admit the same
number of isometries as flat Minkowski space-time).
Our real universe is probably not asymptotically flat and
the whole theory of gravitational radiation should eventually be formulated
with other boundary conditions than those corresponding to
asymptotic flatness. Any exact explicit example of a wave propagating in a
space-time which is not asymptotically flat may give a useful insight.
It may also serve as a `test-bed' for numerical simulations.

In the next section we shall review the Siklos class of
solutions. It will be shown to be identical with one subclass
of space-times studied by Ozsv\' ath, Robinson and R\' ozga
\cite{ORR}. In Section 3 we shall analyze the equation of geodesic
deviation in  frames
parallelly transported along time-like geodesics. It will be demonstrated
that this choice (although being the most natural one)
is not suitable for a physical interpretation. The simple
interpretation of the vacuum Siklos space-times (here by vacuum
space-times we understand Einstein spaces with $\Lambda<0$),
presented in Section 4,
will be  given in rotating frames in which the solutions
clearly represent exact gravitational waves
propagating `in' the anti-de Sitter universe. A
surprising result is that the direction, in which the
waves propagate, rotates with angular velocity
$\omega=\sqrt{-\Lambda/3}\>$.
In Section 5 the Kaigorodov solution \cite{Kaig}, an interesting
representative of the Siklos class, will be described
(it is a homogeneous type N vacuum solution with $\Lambda<0$
admitting 5 Killing vectors). The explicit form of all geodesics and a
general solution of the equation of geodesic deviation in the Kaigorodov
space-time will be presented in Section 6.  Finally, remarks on
its global structure  will be given in Section 7.

\section {The Siklos space-times}

In 1985 Siklos found an interesting class of type N space-times
for which the quadruple Debever-Penrose null vector field {\bf k} is also a
Killing vector field \cite{Siklos}.
The metric can be written in the form
\begin{equation}
ds^2={{\beta^2}\over{x^2}}(dx^2+dy^2+2dudv+Hdu^2)\ ,
\label{E2.1}
\end{equation}
where $\beta=\sqrt{-3/\Lambda}$, $\Lambda$ is a negative
cosmological constant, $x$ and $y$ are spatial coordinates,
$v$ is the affine parameter along the rays generated by
${\bf k}=\partial_v$, and $u$ is the retarded time.
If the vacuum equations (with $\Lambda<0$) are
to be satisfied, the function $H(x,y,u)$ must obey
\begin{equation}
{H_{,xx}-{2\over x}H_{,x}+H_{,yy}}=0\ .
\label{E2.2a}
\end{equation}
An explicit solution to this equation reads \cite{Siklos}
\begin{equation}
H=x^2{\partial\over\partial x}\left( {{f+\bar f}\over x}\right)\equiv
  {1\over2}(f_{,\zeta}+\bar f_{,\bar\zeta})(\zeta+\bar\zeta)
  -(f+\bar f)\ ,
\label{E2.2b}
\end{equation}
where $\zeta\equiv x+iy$ and $f(\zeta,u)$ is an arbitrary function, analytic
in $\zeta$. From (\ref{E2.1}) it is clear that the Siklos
space-times are conformal to {\it pp} -waves. In fact, it was
demonstrated in \cite{Siklos} that they represent the only non-trivial Einstein spaces
conformal to non-flat {\it pp} -waves.

The metric (\ref{E2.1}) represents the anti-de Sitter solution when $H=0$.
Any particular solution of (\ref{E2.2a}) can have an
arbitrary profile $h(u)$. Therefore, as for {\it pp} -waves, sandwich waves
can be obtained by taking $h$ nonzero for a finite period of
retarded time. In particular, by taking $h$ to be a delta
function, impulsive waves can be constructed \cite{PG}.

Note that the Siklos class is identical with the special subclass $(IV)_0$
of non-twisting, non-expanding and shear-free
space-times of the Kundt type found by Ozsv\' ath, Robinson and R\' ozga
\cite{ORR} in the form
\begin{equation}
ds^2=2\frac{1}{p^2} d\xi d\bar\xi -  2\frac{q^2}{p^2} dUdV
 -\frac{q}{p} \>\tilde H(\xi,\bar\xi, U)\, dU^2\ ,
\label{E2.2c}
\end{equation}
where $p=1+\la\xi\bar\xi$ and
$q=(1+\sqrt{-\la}\xi)(1+\sqrt{-\la}\bar\xi)$.
The limit $\Lambda\rightarrow0$ in (\ref{E2.2c}) with $\tilde H$
independent of $\Lambda$ gives
immediately the metric of {\it pp} -waves.
The explicit transformation converting (\ref{E2.2c}) to (\ref{E2.1}) is
\begin{equation}
 \xi={-\sqrt{-{6\over\Lambda}}}{{(x+1/2)+iy}\over{(x-1/2)+iy}}
 \ , \quad
 U ={1\over\Lambda}\sqrt{3\over2}u\ , \quad
 V ={12\sqrt{2\over3}}v\ ,
\label{E2.2d}
\end{equation}
so that $\tilde H=-4\Lambda H/x$.
Before concluding this brief introductory section, let us present
the Christoffel symbols for the metric (\ref{E2.1}) in coordinates
$x^\mu=(v,x,y,u)$,
\begin{eqnarray}
&&\Gamma^0_{01}=-\frac{1}{x}\ ,
      \qquad\Gamma^0_{13}=\frac{1}{2}\,H_{,x}\ ,
      \qquad\Gamma^0_{23}=\frac{1}{2}\,H_{,y}\ ,
      \qquad\Gamma^0_{33}=\frac{1}{2}\,H_{,u}\ ,\nonumber\\
&&\Gamma^1_{03}=\ \> \frac{1}{x}\ ,
      \qquad\>\Gamma^1_{11}=-\frac{1}{x}\ ,
      \qquad\ \ \>\Gamma^1_{22}=\ \frac{1}{x}\ ,
      \qquad\quad\>\Gamma^1_{33}=\frac{H}{ x}-\frac{1}{2}\,H_{,x}\ ,\label{E2.3}\\
&&\Gamma^2_{12}=-\frac{1}{x}\ ,
      \qquad\Gamma^2_{33}=-\frac{1}{2}\,H_{,y}\ , \qquad
\Gamma^3_{13}=-\frac{1}{x}\ ,\nonumber
\end{eqnarray}
and all independent nonvanishing components of the Riemann tensor and the
Weyl tensor,
\begin{eqnarray}
&&R_{1013}=R_{2023}=R_{3003}=R_{1212}=F\ , \nonumber\\
&&R_{1313}={1\over2}F(2H-xH_{,x}+x^2H_{,xx})\ ,\quad
  R_{2323}={1\over2}F(2H-xH_{,x}+x^2H_{,yy})\ , \label{E2.4a}\\
&&R_{1323}={1\over2}Fx^2H_{,xy}=C_{1323}\ ,\quad
  C_{1313}=-C_{2323}={1\over4}Fx^2(H_{,xx}-H_{,yy})\ ,\nonumber
\end{eqnarray}
where $F=-{\beta^2}/{x^4}$.

\section {Particles in the Siklos space-times}

In this section we derive an  invariant form of the equation
of geodesic deviation for the Siklos space-times which will be
used in the next section for a physical interpretation.
We consider an arbitrary  test particle freely falling
along timelike geodesic $x^\alpha(\tau)$
with $\tau$ being a proper time normalizing the particle four-velocity
$u^\alpha=dx^\alpha/d\tau$ so that
\begin{equation}
u_\alpha u^\alpha = \epsilon\ ,
\label{E2.6}
\end{equation}
where $\epsilon=-1$ (for $\epsilon=0$ and $\epsilon=+1$ the
geodesic would be null or spacelike, respectively).
For the metric (\ref{E2.1}) the geodesic equations and equation
(\ref{E2.6}) are
\begin{eqnarray}
\ddot v &=& 2\dot v {\dot x\over x} -Cx^2(\dot xH_{,x}+\dot yH_{,y})
             -{1\over2}C^2x^4H_{,u}\ ,     \nonumber\\
{\left({\dot x\over x}\right)^\cdot} &=&-[2C\dot v+ {\dot y^2\over x^2}
           +C^2x^2(H-{1\over2}xH_{,x})]\ , \label{E2.8}\\
\ddot y &=& 2\dot y {\dot x\over x} +{1\over2}C^2x^4H_{,y}
            \ ,\qquad
\dot u  \ =\  Cx^2\ , \nonumber\\
\left({\dot x\over x}\right)^2 &=& -(2C\dot v + {{\dot y}^2\over
            {x^2}} + C^2x^2H-{\epsilon\over {\beta^2}})\ , \nonumber
\end{eqnarray}
where $\cdot\equiv d/d\tau$ and $C=const$
($C\not=0$ for $\epsilon=-1$ since otherwise $\dot u=0$
would be in contradiction with (\ref{E2.6})).
Some particular solutions for a special form of $H$
will be presented later on.
However,  for our purposes, it is not necessary to solve
(\ref{E2.8}) explicitly. Our analysis will primarily be
based on the equation of geodesic deviation
\begin{equation}
{{D^2Z^\mu}\over{d\tau^2}}=-R^\mu_{\alpha\beta\gamma}
    u^\alpha Z^\beta u^\gamma\ ,
\label{E2.9}
\end{equation}
an equation for a displacement vector $Z^\mu (\tau)$
connecting two neighbouring free test particles. In order to obtain an
{\it invariant} relative motion
we set up  an orthonormal frame
$\{ {\bf e}_a \}=\{ {\bf u},{\bf e}_{(1)},{\bf e}_{(2)},{\bf
e}_{(3)}\}$,
${\bf e}_a \cdot {\bf e}_b\equiv g_{\alpha\beta}
                   e_a^\alpha e_b^\beta=\eta_{ab}$.
By projecting (\ref{E2.9}) onto the frame we get
\begin{equation}
\ddot Z^{(i)}=-R^{(i)}_{\ (0)(j)(0)}Z^{(j)}\ ,
\label{E2.11}
\end{equation}
where $Z^{(i)}\equiv e^{(i)}_\mu Z^\mu$ are frame components
of the displacement vector and
$\ddot Z^{(i)}\equiv e^{(i)}_\mu {{D^2Z^\mu}\over{d\tau^2}}$ are
relative accelerations.
We start with a natural choice, namely a frame $\{ {\bf e}_a(\tau) \}$
given by
\begin{eqnarray}
e^\alpha_{(0)}&=&u^\alpha=(\dot v,\dot x,\dot y,Cx^2)\ , \nonumber\\
e^\alpha_{(1)}&=&\sin\tb \big(\dot v+{1\over{\beta^2C}}\;,\dot x,\dot y,Cx^2\big)
             -\cos \tb \big({1\over{\beta C}}{\dot x\over x}\;,-{x\over\beta},0,0\big)
             \ , \nonumber\\
e^\alpha_{(2)}&=&{x\over\beta}\big(-{\dot y\over{Cx^2}}\;,0,1,0\big)\ ,\label{E2.14}\\
e^\alpha_{(3)}&=&\cos\tb \big(\dot v+{1\over{\beta^2C}}\;,\dot x,\dot y,Cx^2\big)
             +\sin\tb \big({1\over{\beta C}}{\dot x\over x}\;,-{x\over\beta},0,0\big)\ ,
\nonumber
\end{eqnarray}
that is {\it parallelly transported} along any timelike
geodesic in the Siklos spacetime.
Next step is to calculate the frame components of the Riemann tensor
by using (\ref{E2.4a}) and (\ref{E2.14}). Straightforward but somewhat
tedious calculations give
\begin{eqnarray}
&&R_{(1)(0)(1)(0)}=-\frac{\Lambda}{3}+\A_+\cos^2\tb\ , \hskip20mm
 R_{(2)(3)(2)(3)}=\ \>\frac{\Lambda}{3}-\M\cos^2\tb\ , \nonumber\\
&&R_{(3)(0)(3)(0)}=-\frac{\Lambda}{3}+\A_+\sin^2\tb\ , \hskip21mm
 R_{(1)(2)(1)(2)}=\ \>\frac{\Lambda}{3}+\M\sin^2\tb\ ,\nonumber\\
&&R_{(1)(3)(1)(3)}=\ \>\frac{\Lambda}{3}+\A_+\ , \hskip39mm
 R_{(2)(0)(2)(0)}=-\frac{\Lambda}{3}-\M\ ,  \nonumber\\
&&R_{(1)(0)(2)(0)}=R_{(1)(3)(2)(3)}=- \A_\times\cos\tb\ , \nonumber\\
&&R_{(2)(0)(3)(0)}=R_{(1)(2)(1)(3)}=\ \>\A_\times\sin\tb\ ,\label{E2.14a}\\
&&R_{(0)(1)(1)(3)}=-\A_+\cos\tb\ , \hskip31mm
 R_{(0)(2)(2)(3)}=\ \M\cos\tb\ , \nonumber\\
&&R_{(0)(3)(1)(3)}=\ \A_+\sin\tb\ , \hskip33mm
 R_{(0)(2)(1)(2)}=-\M\sin\tb\ , \nonumber\\
&&R_{(1)(0)(3)(0)}=-\A_+\sin\tb\cos\tb\ , \hskip16.5mm
 R_{(1)(2)(2)(3)}=\ \M\sin\tb\cos\tb\ , \nonumber\\
&&R_{(0)(1)(1)(2)}=R_{(0)(3)(2)(3)}=- \A_\times\sin\tb\cos\tb\ , \nonumber\\
&&R_{(0)(1)(2)(3)}=\A_\times\cos^2\tb\ , \quad\
  R_{(0)(3)(1)(2)}=\A_\times\sin^2\tb\ , \quad\
  R_{(0)(2)(1)(3)}=\A_\times\ , \nonumber
\end{eqnarray}
where
\begin{equation}
\A_+=-\frac{1}{2}C^2x^5\left(\frac{H_{,x}}{x}\right)_{,x}\ ,\quad
\A_\times=\frac{1}{2}C^2x^5\left(\frac{H_{,x}}{x}\right)_{,y}\ ,\quad
\M=\frac{1}{2}C^2x^3(xH_{,yy}-H_{,x})\ .   \label{E2.14b}
\end{equation}
By substituting the components (\ref{E2.14a}) into (\ref{E2.11}) we get
\begin{eqnarray}
\ddot Z^{(1)}&=&{\frac{\Lambda}{3}}Z^{(1)}-\,\A_+\cos\tb
      \Big[\cos\tb Z^{(1)}-\sin\tb Z^{(3)}\Big]
     +\A_\times\cos\tb Z^{(2)}\ ,\nonumber\\
\ddot Z^{(2)}&=&{\frac{\Lambda}{3}}Z^{(2)}+\M Z^{(2)}
     +\A_\times\Big[\cos\tb Z^{(1)}-\sin\tb Z^{(3)}\Big]\ ,\label{E2.15}\\
\ddot Z^{(3)}&=&{\frac{\Lambda}{3}}Z^{(3)}\>+\A_+\sin\tb
      \Big[\cos\tb Z^{(1)}-\sin\tb Z^{(3)}\Big]
     -\A_\times\sin\tb Z^{(2)}\ . \nonumber
\end{eqnarray}
The structure of the equations is not simple. It may seem somewhat surprising
because the Siklos solution is of Petrov type N so that it should describe
gravitational waves affecting motions only in
directions perpendicular to the direction of propagation, cf. \cite{Szek}.
However, equations (\ref{E2.15}) can be
simplified with transverse effects becoming evident by a transformation
from (\ref{E2.14}) to another frame. The idea follows naturally
from the components of the quadruple Debever-Penrose vector
${\bf k}=\partial_v$, $k^{(1)}=\beta^2 C\sin\tb$, $k^{(2)}=0$,
$k^{(3)}=\beta^2 C\cos\tb$,
which indicate that the {\it spacelike direction of propagation of the
wave  rotates uniformly in the $({\bf e}_{(1)},{\bf e}_{(3)})$ plane}.
Thus, we can define a new frame
$\{ {\bf e}_{a'} \}=\{ {\bf u},{\bf e}_{(1')},{\bf e}_{(2)},{\bf e}_{(3')}\}$
by
\begin{eqnarray}
&&e^\alpha_{(1')}=\cos\tb e^\alpha_{(1)}-\sin\tb e^\alpha_{(3)}
  =\big(-{1\over{\beta C}}{\dot x\over x} \;,{x\over\beta},0,0\big)\ , \nonumber\\
&&e^\alpha_{(3')}=\sin\tb e^\alpha_{(1)}+\cos\tb e^\alpha_{(3)}
  =\big(\dot v+{1\over{\beta^2C}}\;,\dot x, \dot y,Cx^2\big) \ , \label{E2.17}
\end{eqnarray}
in which the vector ${\bf k}$ has components $k^{(1')}=0=k^{(2)}$,
$k^{(3')}=\beta^2 C \not =0$.
The orthonormal frame $\{ {\bf e}_{a'} \}$ {\it is not parallelly
transported} along any timelike geodesic since it
rotates uniformly  with respect to (\ref{E2.14}).
Using (\ref{E2.17}) we can rewrite (\ref{E2.15}) as
\begin{eqnarray}
\ddot Z^{(1')}&=&{\frac{\Lambda}{3}}Z^{(1')}-\A _+ Z^{(1')}
    +\A _\times Z^{(2)}\ , \nonumber\\
\ddot Z^{(2)}\,&=&{\frac{\Lambda}{3}}Z^{(2)}\>+\M\> Z^{(2)}\>
    +\A _\times Z^{(1')}\ , \label{E3.19b}\\
\ddot Z^{(3')}&=&{\frac{\Lambda}{3}}Z^{(3')}\ . \nonumber
\end{eqnarray}
This can be used for the interpretation of {\it general}
Siklos space-times. In the following however,  we concentrate
only on vacuum solutions (with $\Lambda<0$) describing
`pure' gravitational waves in the absence of matter.

\section {Vacuum Siklos space-times as exact gravitational waves
          in the anti-de Sitter universe}

Using the field equation (\ref{E2.2a}) and its solution
(\ref{E2.2b}) we get
\begin{eqnarray}
&&\A _+(\tau)= -\frac{1}{2}C^2x^5\left({{H_{,x}}\over x}\right)_{,x}
  \equiv -{C^2\over 32}(\zeta+\bar\zeta)^5\,{\cal R}
  e\{f_{,\zeta\zeta\zeta}\}=\M\ , \nonumber\\
&&\A _\times(\tau)=+\frac{1}{2}C^2x^5\left({{H_{,x}}\over x}\right)_{,y}
  \equiv -{C^2\over 32}(\zeta+\bar\zeta)^5\,{\cal I}
  m\{f_{,\zeta\zeta\zeta}\}\ .
\label{E3.21}
\end{eqnarray}
The system (\ref{E3.19b}) with (\ref{E3.21})
represents the main result of our analysis.
It is particularly well  suited for the physical interpretation of
vacuum Siklos space-times:

\begin{itemize}
\item [1.] All test particles move isotropically one with respect to the other
$\Big( \ddot Z^{(i)}=\frac{\Lambda}{3}Z^{(i)}\> ,$ $i=1',2,3'\Big)$
if  $\A _+=0=\A _\times$, i.e. if
$H(x,y,u)=c_0(u)+c_1(u)y+c_2(u)\left(x^2+y^2\right)$
corresponding to $f_{,\zeta\zeta\zeta}=0$.
No gravitational wave is present in this case.
This agrees with the fact that for $H$ of this form  the Siklos
solution is conformally flat --- the Weyl tensor vanishes
(see (\ref{E2.4a})).
The only conformally flat vacuum solution with $\Lambda<0$ is the
anti-de Sitter spacetime, maximally symmetric
solution of constant negative curvature. This explains the resulting
isotropic motions. Thus, the terms
proportional to $\Lambda$ in (\ref{E3.19b}) represent the influence of the
{\it anti-de Sitter background}.

\item [2.] If the amplitudes $\A _+$ and $\A _\times$ do not vanish
(which is for  $f_{,\zeta\zeta\zeta}\not=0$) the
particles  are influenced (for $\Lambda \rightarrow 0$) similarly
as by standard  gravitational waves on Minkowski background
(such as exact {\it pp} -waves or linearized waves \cite{MTW}). However, for
$\Lambda<0$ the influence of the gravitational wave adds with the anti-de
Sitter isotropic background motions due to the presence of the
$\Lambda$-terms. This supports our interpretation of the Siklos solution as
an {\it exact gravitational wave in the anti-de Sitter universe}.

\item [3.] The gravitational {\it wave propagates in the spacelike direction of}
${\bf e}_{(3')}$ and has a {\it transverse character} since only motions in the
perpendicular directions ${\bf e}_{(1')}$ and ${\bf e}_{(2)}$ are affected.
The direction of propagation is {\it not} parallelly transported
--- it uniformly rotates with respect to parallel frames
along any geodesic with angular velocity given by
$\omega=\sqrt{-\Lambda/3}$. In the limit $\Lambda\rightarrow0$
the effect of rotation vanishes.

\item [4.] The wave has {\it two polarization modes}: `$+$' and
`$\times$' with $\A _+$ and $\A _\times$ being the corresponding two
independent {\it amplitudes}. The amplitudes given by (\ref{E3.21}) depend on
the proper time of each particle through $x(\tau)$ and
$H(x(\tau),y(\tau),u(\tau))$ where $x^\mu (\tau)$ describes the geodesic.
Performing the rotation in the transverse plane
\begin{equation}
\tilde {\bf e}_{(1')}=\cos\vartheta\> {\bf e}_{(1')} +
    \sin\vartheta\> {\bf e}_{(2)}\ ,\quad
\tilde {\bf e}_{(2)}\>=-\sin\vartheta\> {\bf e}_{(1')} +
    \cos\vartheta\> {\bf e}_{(2)}\ ,
\label{E3.22}
\end{equation}
the motions are again given as in (\ref{E3.19b}), only the
amplitudes change according to
\begin{equation}
\tilde\A _+(\tau)= \cos 2\vartheta\> \A_+ -\sin 2\vartheta\>\A_\times\ ,
\quad
\tilde\A _\times(\tau)= \sin 2\vartheta\> \A_+ +\cos
2\vartheta\>\A_\times\ .
\label{E3.23}
\end{equation}
Relations (\ref{E3.23}) represent the transformation
(polarization) properties of the wave amplitudes.
They are $\pi$-periodic so that the helicity is equal to 2.
Moreover, by special choices of the polarization parameter $\vartheta=
\vartheta_+$ or $\vartheta=\vartheta_\times$ one can set up at any event
privileged frames in which either $\tilde\A_\times=0$ or $\tilde\A_+=0$,
i.e., the wave is purely polarized.
Since $\vartheta_\times =\vartheta_+ +{\pi\over4}$, the two modes are
${\pi\over4}$ - shifted.

\item [5.] From (\ref{E3.21}) it follows that radiative vacuum Siklos
space-times contain singularities at $\zeta+\bar\zeta=\infty$ (corresponding
to $x=\infty$) since the components (\ref{E2.14a}) of the Riemann
tensor are proportional to diverging gravitational-wave
amplitudes. According to definitions presented in
\cite{ES, HE} there is a
curvature singularity at $x=\infty$. Other singularities
arise if $f_{,\zeta\zeta\zeta}$ in the amplitudes diverges.
\end{itemize}

We conclude this section by rewriting the equation of geodesic
deviation. The form (\ref{E3.19b})  is well suited for
interpretation due to its simple structure but it is not
useful for looking for solutions:
$\ddot Z^{(i)}=e^{(i)}_\mu ({D^2Z^\mu}/{d\tau^2})$
is not a total time derivative of $Z^{(i)}(\tau)$ for $i=1',3'$ since
${\bf e}_{(1')},{\bf e}_{(3')}$ are not parallelly transported. In fact,
\begin{equation}
{{d^2Z^{(i)}(\tau)}\over{d\tau^2}}=\ddot Z^{(i)}+
               2{{De^{(i)}_\mu}\over{d\tau}}{{DZ^\mu}\over{d\tau}}+
               Z^\mu{{D^2e^{(i)}_\mu}\over{d\tau^2}}\ .
\label{E3.25}
\end{equation}
For (\ref{E2.17}) we get by using of
${{D{\bf e}_a}/{d\tau}}=0$
\begin{equation}
{{D{\bf e}^{(1')}}\over{d\tau}}=-{1\over\beta}{\bf e}^{(3')}\ ,\
{{D{\bf e}^{(3')}}\over{d\tau}}\>=\>{1\over\beta}{\bf e}^{(1')}\ ,\
{{D^2{\bf e}^{(1')}}\over{d\tau^2}}=-{1\over{\beta^2}}{\bf e}^{(1')}
\ ,\
{{D^2{\bf e}^{(3')}}\over{d\tau^2}}=-{1\over{\beta^2}}{\bf e}^{(3')}
\ .
\label{E3.26}
\end{equation}
Equations (\ref{E3.25}) thus take the form
\begin{equation}
\ddot Z^{(1')}={{d^2Z^{(1')}}\over{d\tau^2}}+
    {2\over\beta}{{dZ^{(3')}}\over{d\tau}}-{1\over{\beta^2}}Z^{(1')}
    \ ,\qquad
\ddot Z^{(3')}={{d^2Z^{(3')}}\over{d\tau^2}}-
    {2\over\beta}{{dZ^{(1')}}\over{d\tau}}-{1\over{\beta^2}}Z^{(3')}
    \ .
\label{E3.27}
\end{equation}
By combining (\ref{E3.27}) with (\ref{E3.19b}) we get the following
form of the equation of geodesic deviation
\begin{eqnarray}
{{d^2Z^{(1')}}\over{d\tau^2}}+\left({{4\over{\beta^2}}+\A_+(\tau)}\right)
    Z^{(1')} &=& \A_\times(\tau)Z^{(2)}-{2\over\beta}C_1\ , \nonumber\\
{{d^2Z^{(2)}}\over{d\tau^2}}\;+\left({{1\over{\beta^2}}-\A_+(\tau)}\right)
    Z^{(2)}\>&=& \A_\times(\tau)Z^{(1')}\ , \label{E3.28}\\
Z^{(3')}&=&{2\over\beta}\int Z^{(1')}\, d\tau +C_1\tau+C_2\ ,
\nonumber
\end{eqnarray}
$C_1,C_2$ being constants. The system can be integrated
provided we know the  explicit form of the geodesic
$x^\mu(\tau)$ and therefore of the amplitudes $\A_+(\tau)$,
$\A_\times(\tau)$. (In Section 6 we shall present a general
solution of (\ref{E3.28}) for the case when $H=x^3$.)
Let us note here only that there always exists a trivial solution
(along {\it any} timelike geodesic in {\it any}
vacuum Siklos space-time) given by
$Z^{(1')}=0=Z^{(2)}$, $Z^{(3')}=D$, $D$ being a constant, i.e.
(cf. (\ref{E2.17})),
\begin{equation}
Z^{(1)}\;=D\sin\tb,\quad Z^{(2)}=0,\quad Z^{(3)}\;=D\cos\tb\ .
\label{E3.29}
\end{equation}
It has a simple interpretation: the particles may always
{\it corotate uniformly in circles} with constant
angular velocity $\omega=\sqrt{-{\Lambda}/{3}}$ around the `fiducial'
reference particle (if measured with respect to parallelly transported frames).
For $\Lambda\rightarrow 0$ the rotation vanishes.

\section {The Kaigorodov space-time}

As an interesting particular example of the Siklos-type metric (1) we now
analyze a vacuum solution with $\Lambda<0$ given by $H=x^3$ corresponding to
$f={1\over4}\zeta^3$ (cf. (\ref{E2.2b})),
\begin{equation}
ds^2={\beta^2\over x^2}(dx^2+dy^2+2dudv+x^3 du^2) \ .
\label{E4.0}
\end{equation}
In fact, such a solution represents the simplest non-trivial vacuum
space-time of the Siklos type (for $f$ quadratic in $\zeta$ one gets just
the conformally flat anti-de Sitter space-time, cf. (\ref{E3.21})).
Therefore, it can be understood as a $\Lambda<0$
analogue of the ``homogeneous'' $pp$ -wave in Minkowski background
\cite{EK} which is also the simplest vacuum $pp$ space-time.
The solution (\ref{E4.0})  was first discovered by Kaigorodov
\cite{Kaig} in the form
\begin{equation}
ds^2=(dx^4)^2+e^{2{{x^4}/\beta}}[2dx^1dx^3+(dx^2)^2]
        \pm e^{-{{x^4}/\beta}}(dx^3)^2\ .
\label{E4.1}
\end{equation}
Transformation between the Kaigorodov and the Siklos coordinates is given by
\begin{equation}
x^1=\beta v \ ,\quad
x^2=\beta y \ ,\quad
x^3=\beta u \ ,\quad
x^4=-\beta\ln |x| \ .
\label{E4.3}
\end{equation}
The solution has also been discussed independently in
\cite{Cahen}-\cite{Ozs2} and it is
a special case of the $(IV)_0$ class found by
Ozsv\' ath, Robinson and R\' ozga (see (\ref{E2.2c}) and
Eq. (6.17) in \cite{ORR}),
\begin{equation}
ds^2=2\frac{1}{p^2} d\xi d\bar\xi -  2\frac{q^2}{p^2} dUdV
 +\Lambda\frac{p}{q}\, dU^2\ ,
\label{EE4}
\end{equation}
the transformation to (\ref{E4.0}) being given by (\ref{E2.2d}).
Other forms of the Kaigorodov solution can be found in
\cite{KSMH}, Eq. (10.33),
\begin{equation}
ds^2=-{12\over\Lambda}dZ^2+10ke^{2Z}dX^2+e^{-4Z}dy^2-10Ue^ZdZdX-2e^ZdUdX
\ ,
\label{EE6}
\end{equation}
resulting from the transformation
\begin{equation}
x=\beta e^{2Z}  \ ,\quad
u=-\sqrt{{10k}\over{\beta^3}} X \ , \quad
v=\sqrt{{\beta^3}\over{10k}} e^{5Z}U \ ,
\label{EE7}
\end{equation}
and Eq. (33.2) (there is a misprint in \cite{KSMH}: the coefficient
$2(\Lambda x)^{-2}$  should be $-3/(\Lambda x^2)$),
\begin{equation}
ds^2={{\beta^2}\over{x^2}}(dx^2+dy^2)-2dU(dV+2V{{dx}\over x}-xdU)
\ ,
\label{EE8}
\end{equation}
resulting from
\begin{equation}
u={{\sqrt 2}\over\beta} U \ ,\quad
v=-{1\over{\sqrt{2} \beta}} Vx^2\ .
\label{EE9}
\end{equation}

The Kaigorodov space-time is the only homogeneous type N solution
(the quadruple Debever-Penrose vector being ${\bf k}=\partial_v$) of
the Einstein vacuum field equations with $\Lambda\not =0$  (necessarily
with $\Lambda<0$). It admits a five-parameter group of motions.
The Killing vectors in the Siklos coordinates $(v,x,y,u)$ are
(see \cite{Siklos})
\begin{eqnarray}
{\xi}^\mu_{(1)}&=&(1,0,0,0)     \ ,\qquad
{\xi}^\mu_{(2)}\ =\ (0,0,1,0)   \ ,\qquad
{\xi}^\mu_{(3)}\ =\ (0,0,0,1)   \ ,\nonumber\\
{\xi}^\mu_{(4)}&=&(-y,0,u,0)    \ ,\quad
{\xi}^\mu_{(5)}\ =\ (5v,2x,2y,-u) \ , \label{EE10}
\end{eqnarray}
the corresponding isometries being:
1.~$v'=v+v_0$,
2.~$y'=y+y_0$,
3.~$u'=u+u_0$,
4.~$v'=-{{A^2}\over 2}u-Ay+v$, $y'=Au+y$,
5.~$v'=e^{5B}v$, $x'=e^{2B}x$, $y'=e^{2B}y$, $u'=e^{-B}u$.
From (\ref{EE10}) we see that the quadruple
Debever-Penrose null vector is also a Killing vector.
However, it is not covariantly constant.

\section {Particles in the Kaigorodov space-time}

For $H=x^3$ representing the Kaigorodov solution, the
equations of motion  (\ref{E2.8}) give
\begin{eqnarray}
 \dot x^2&=&C^2x^7-(2AC+B^2)x^4+\epsilon{\frac{x^2}{\beta^2}}
  \ ,\nonumber\\
   \dot v&=& Ax^2-Cx^5 \ ,\qquad
   \dot y\ =\  Bx^2  \ , \qquad \dot u\ =\  Cx^2      \ ,\label{E5.31b}\\
   x\ddot x-\dot x^2&=&{5\over2}C^2x^7-(2AC+B^2)x^4  \ ,\nonumber
\end{eqnarray}
$A,B,C$ being real constants of integration. Now we must
distinguish two cases.

\noindent
{\bf Case 1.} If $\dot x=0$ then the equations (\ref{E5.31b}) give
\begin{eqnarray}
  v(\tau)&=&\left({Ax^2_0-Cx^5_0}\right)\tau+v_0 \ ,\quad
  x(\tau)\ =\ x_0            \ ,\nonumber\\
  y(\tau)&=&Bx^2_0\tau+y_0 \ , \hskip20mm
  u(\tau)\ =\ Cx^2_0\tau+u_0 \ , \label{E5.32}
\end{eqnarray}
where $v_0, x_0, y_0, u_0$ and  $A, B, C$ are real constants satisfying
the conditions ${3\over2}C^2x^5_0\beta^2=\epsilon$ and
${5\over2}C^2x^3_0=2AC+B^2$.
The first condition implies that for $x_0<0$, $C\not=0$ all the geodesics
are timelike $(\epsilon=-1)$  and for $x_0>0$, $C\not=0$ spacelike
$(\epsilon=+1)$. For $C=0$ the geodesics are null $(\epsilon=0)$
and they have a simple form $v=Ax^2_0\tau+v_0$, $x=x_0$, $y=y_0$, $u=u_0$
since the second condition gives $B=0$ in this case.

\noindent
{\bf Case 2.} If $\dot x\not=0$ then the last equation in (\ref{E5.31b})
can simply be omitted (since the first equation is its
integral). The four remaining equations give
\begin{eqnarray}
  \tau-\tau_0&=&\int \Big(1/x\sqrt
            {C^2x^5-(2AC+B^2)x^2+\epsilon/\beta^2}\Big)\>dx\ , \nonumber\\
  v(\tau)&=&A\int x^2(\tau)\,d\tau -C\int x^5(\tau)\,d\tau+v_0\ ,\label{E5.33} \\
  y(\tau)&=&B\int x^2(\tau)\,d\tau+y_0\ ,   \qquad
  u(\tau)\ =\ C\int x^2(\tau)\,d\tau+u_0\ ,  \nonumber
\end{eqnarray}
where $\tau_0, v_0, y_0, u_0, A, B, C$ are arbitrary constants.
For special values of the parameters the integrations can be
performed analytically:
\begin{itemize}
\item [(i)] If $C=0$ then the geodesics must be
spacelike. Their form is given either by
\begin{eqnarray}
  v(\tau)&=&A\frac{\beta}{2} \exp{[2(\tau-\tau_0)/\beta]}+v_0 \ ,\nonumber \\
  x(\tau)&=&\pm \exp{[(\tau-\tau_0)/\beta]}  \ ,\label{E5.33a} \\
  y(\tau)&=&y_0 \ , \quad  u(\tau)\ =\ u_0 \ ,\nonumber
\end{eqnarray}
(for $B=0$), or by
\begin{eqnarray}
  v(\tau)&=&\frac{A}{\beta B^2}\tanh{[(\tau-\tau_0)/\beta]}+v_0 \ ,\nonumber \\
  x(\tau)&=&\pm\left(\beta |B|\cosh{[(\tau-\tau_0)/\beta]}\right)^{-1}  \ ,\label{E5.33b} \\
  y(\tau)&=&\frac{1}{\beta B}\tanh{[(\tau-\tau_0)/\beta]}+y_0 \ , \quad
  u(\tau)=u_0 \ ,\nonumber
\end{eqnarray}
(for $B\not=0$).
\item [(ii)] If $C\not=0$ it is convenient to
further simplify (\ref{E5.33}) by using the symmetries of the
solution. For a Killing vector
${\xi}^\mu$, the expression $u_\mu \xi^\mu$ is the constant of
motion for any geodesic observer having the four-velocity
$u^\mu$. The Killing vectors ${\xi}^\mu_{(1)}$,
${\xi}^\mu_{(2)}$ and ${\xi}^\mu_{(3)}$ (see (\ref{EE10}))
give relations embodied already in
(\ref{E5.31b}), the vectors ${\xi}^\mu_{(4)}$ and ${\xi}^\mu_{(5)}$
imply $Bu-Cy=const$ and
$\dot x/x+\frac{5}{2}Cv+By-\frac{1}{2}Au=const$,
respectively. These two relations simplify (\ref{E5.33}) into
\begin{eqnarray}
  \tau-\tau_0&=&\int \Big(1/x\sqrt
            {C^2x^5-(2AC+B^2)x^2+\epsilon/\beta^2}\Big)\>dx\ , \nonumber\\
  u(\tau)&=&C\int x^2(\tau)\,d\tau+u_0\ ,\quad
  y(\tau)\>=\>\frac{B}{C} u(\tau)+y_0\ ,  \label{E5.33e} \\
v(\tau)&=&\frac{2}{5C}\left[\left(\frac{A}{2}-\frac{B^2}{C}\right)u(\tau)
        -\frac{\dot x(\tau)}{x(\tau)}\right]+v_0\nonumber
\end{eqnarray}
(we have reparametrized the constants $y_0$ and $v_0$). In the case
when $2AC+B^2=0$, the remaining two integrations can be performed
explicitly:
{\it null} geodesics are
\begin{eqnarray}
  v(\tau)&=&\frac{4}{25C}{(\tau-\tau_0)}^{-1}
           +\frac{B^2}{C|C|}{\left[{-\frac{5}{2}|C|(\tau-\tau_0)}\right]}^{1/5}
           -\frac{B^2}{2C^2}u_0+v_0 \ ,\nonumber \\
  x(\tau)&=&{\left[{-\frac{5}{2}|C|(\tau-\tau_0)}\right]}^{-2/5}  \ ,\nonumber \\
  y(\tau)&=&-\frac{2B}{|C|}{\left[{-\frac{5}{2}|C|(\tau-\tau_0)}\right]}^{1/5}
           +\frac{B}{C}u_0+y_0 \ ,\label{E5.33f} \\
u(\tau)&=&-\frac{2C}{|C|}{\left[{-\frac{5}{2}|C|(\tau-\tau_0)}\right]}^{1/5}
           +u_0 \ .\nonumber
\end{eqnarray}
{\it Timelike} geodesics are
\begin{eqnarray}
  v(\tau)&=&-{2\over{5\beta C}}\tan t\
          -\frac{B^2}{2C^2}u(\tau)+v_0\ , \nonumber \\
  x(\tau)&=&(\beta C\cos t)^{-2/5}\ , \quad
  y(\tau)\ =\ \frac{B}{C}u(\tau)+y_0\ ,  \label{E5.34} \\
  u(\tau)&=&-2(\beta C\cos t)^{1/5}C^{\left ({3/5}\right )}_{-1}
           (\sin t)+u_0\ , \nonumber
\end{eqnarray}
where $t=5\tau/2\beta+\tau_0$ and the symbol $C^{(b)}_n(z)$
denotes the Gegenbauer polynomial (generalized for negative values of $n$)
which is a particular case of the hypergeometric function,
$C^{(b)}_n(z)\equiv N^{(b)}_nF(-n,n+2b; b+1/2;(1-z)/ 2)$.
(For $n>0$ the standard normalization
coefficient is $N^{(b)}_n=\Gamma (n+2b)/(n!\Gamma(2b))$;
in order to avoid singularities we define $N^{(b)}_n=1$ for
$n<0$. It can be shown that for {\it non-integer} $a$,
\begin{equation}
\int_0^t \cos^a t\, dt={1\over 1+a}\left(
 C^{\left (1+a/2\right )}_{-1}(0)-
 \cos ^{1+a} t\, C^{\left (1+a/2\right )}_{-1}
 (\sin t)\right)\ ,
\label{E5.35}
\end{equation}
which we have used in (\ref{E5.34}) for $a=-4/5$).
Note that for $A=0=B$, the geodesics (\ref{E5.34}) are highly
{\it privileged} since they are perpendicular  $(u_\mu \xi^\mu =0)$ to
the Killing vectors {\boldmath $\xi$}$\,_{(2)}$, {\boldmath
$\xi$}$\,_{(3)}$. Moreover, for $y_0=0=v_0$ they are perpendicular
to {\boldmath $\xi$}$\,_{(4)}$, {\boldmath $\xi$}$\,_{(5)}$, too.

Finally, {\it spacelike} geodesics (\ref{E5.33e}) for $2AC+B^2=0$
take the form
\begin{eqnarray}
  v(\tau)&=&{2\over{5\beta C}}\coth t
          -\frac{B^2}{2C^2}u(\tau)+v_0\ , \nonumber \\
  x(\tau)&=&(\beta C\sinh t)^{-2/5}\ , \quad
  y(\tau)\>=\>\frac{B}{C}u(\tau)+y_0\ ,  \label{E5.35a} \\
  u(\tau)&=&\frac{2}{5}(\beta C)^{1/5}  \int \sinh^{-4/5} t\,dt
           +u_0\ . \nonumber
\end{eqnarray}
All the geodesics (\ref{E5.33f}), (\ref{E5.34}) and (\ref{E5.35a})
are in the region $x>0$. There are no null and timelike geodesics
of the form (\ref{E5.33e}) with $2AC+B^2=0$ in the region $x<0$;
there are only spacelike geodesics
\begin{eqnarray}
  v(\tau)&=&{2\over{5\beta C}}\tanh t
          -\frac{B^2}{2C^2}u(\tau)+v_0\ , \nonumber \\
  x(\tau)&=&-(-\beta C\cosh t)^{-2/5}\ , \quad
  y(\tau)\ =\ \frac{B}{C}u(\tau)+y_0\ ,  \label{E5.35b} \\
  u(\tau)&=&-\frac{2}{5}(-\beta C)^{1/5}  \int \cosh^{-4/5} t\,dt
           +u_0\ . \nonumber
\end{eqnarray}
\end{itemize}

\noindent
We end this section by investigating the {\it relative} motion of
particles in the Kaigorodov space-time.
The amplitudes (\ref{E3.21}) are $\A _+= -{3\over2}C^2x^5$,
$\A _\times=0$ where $x\equiv x(\tau)$ depends on the {\it timelike}
geodesic, i.e., it is given by (\ref{E5.32}), (\ref{E5.33e}), or (\ref{E5.34})
in particular. The frame is now privileged since the wave looks purely `+'
polarized. Therefore, (\ref{E3.28}) decouples into
the `Schr\" odinger-type' equations, $\A _+$ being a `potential',
and explicit solutions  can be found:

\noindent
{\bf Case 1.} If $\dot x=0$ then the geodesics are given by (\ref{E5.32})
with $\epsilon=-1$ implying $\A _+= -{3\over2}C^2x^5_0=1/\beta^2$.
In this case the general solution of (\ref{E3.28}) is
\begin{eqnarray}
Z^{(1')} &=&D\cos \Big ({{\sqrt 5}\over\beta}\tau+\tau_0\Big )-
          {2\over5}\beta C_1\ , \nonumber \\
Z^{(2)}\,&=&E_1\tau+E_2\ ,   \label{E5.38}\\
Z^{(3')} &=&{2\over{\sqrt 5}}D\sin \Big ({{\sqrt 5}\over\beta}\tau+\tau_0\Big )+
          {1\over5}C_1\tau+C_2\ , \nonumber
\end{eqnarray}
$C_1,C_2,D,E_1,E_2$ being real constants.
The test particles move with a constant velocity $E_1$ in the
direction of ${\bf e}_{(2)}$ whereas in the $({\bf e}_{(1')},{\bf e}_{(3')})$
plane the particles move in {\it ellipses}. In particular, for
$C_1=D=E_1=E_2=0$ we get the solution (\ref{E3.29}) describing uniform
rotations.

\noindent
{\bf Case 2.} If $\dot x\not=0$ then the timelike geodesics are given by
(\ref{E5.33e}) with $\epsilon=-1$. Using (\ref{E5.31b}) one can  verify
that the solution of (\ref{E3.28}) is
\begin{eqnarray}
Z^{(1')} &=&{{\dot x}\over {x^3}}\Big ({D_1+\int {{x^4}\over{\dot x ^2}}
           (D_2x^2+{{C_1}\over\beta})\, d\tau}\Big )\ , \nonumber \\
Z^{(2)}\,&=&\>{1\over x}\>\Big ({E_1+E_2 \int x^2\,d\tau}\Big )
           \ , \label{E5.39}\\
Z^{(3')} &=&{2\over\beta}\int Z^{(1')}\, d\tau +C_1\tau+C_2\ ,  \nonumber
\end{eqnarray}
where $x\equiv x(\tau)$ follows from (\ref{E5.33e}) and is unique for any
geodesic. Since (\ref{E5.39}) contains 6 independent constants of integration
it is a {\it general solution} of the
equation of geodesic deviation. As in the previous case, for
$C_1=D_1=D_2=E_1=E_2=0$ we get the solution (\ref{E3.29}).
In particular, for the geodesics (\ref{E5.34}) with $A=0=B$ we have
$x(\tau)=(\beta C\cos t)^{-2/5}$ for which (\ref{E5.39})
can be integrated into
\begin{eqnarray}
Z^{(1')} &=& D_1\sin t\cos ^{-1/5}t +
   D_2\cos^{6/5}t\; C^{\left (6/5\right )}_{-2}(\sin t) -
   {2\over11}\beta C_1\cos^2 t\;
   C^{\left (8/5\right )}_{-2}(\sin t)\ , \nonumber \\
Z^{(2)}\,&=&E_1\cos ^{2/5}t +E_2\cos^{3/5}t\; C^{\left (3/5\right )}_{-1}
           (\sin t)\ ,  \label{E5.40}\\
Z^{(3')}&=&{4\over5}\int Z^{(1')}\, dt +C_1\tau+C_2  \nonumber
\end{eqnarray}
(we have used (\ref{E5.35}), the identity
$C^{(b)}_{-2}(z)\equiv (2b-1)-(2b-2)zC^{(b)}_{-1}(z)$,
and we have re-defined the constants of integration).
Notice that for $A=0=B$ the amplitude $\A _+=(\Lambda/2)\cos ^{-2} t$ is
independent of the geodesic (given by some value of $C$) and that
$\A _+\rightarrow 0$ as $\Lambda\rightarrow 0$. In such a limit equations
(\ref{E3.28}) reduce to $d^2Z^{(i)}/d\tau^2=0$, $i=1', 2, 3'$.
Considering also
$({\bf e}_{(1')},{\bf e}_{(3')})\rightarrow({\bf e}_{(1)},{\bf e}_{(3)})$
as $\Lambda\rightarrow0$, cf. (\ref{E2.17}), we conclude that
test particles move uniformly as in flat Minkowski space.

\section {On the global structure of the Kaigorodov space-time}

In this section we briefly touch upon some global properties of
the Kaigorodov space-time. The metric (\ref{E4.0})
indicates that the space-time is regular everywhere except at $x=0$
and $x=\infty$. All components of the curvature tensor
for the Kaigorodov solution in the orthonormal frame parallelly
propagated along time-like geodesics are given by
(\ref{E2.14a}) with $\A _+=\M= -{3\over2}C^2x^5$, $\A _\times=0$.
Therefore, $x=0$ represents only a coordinate singularity. However,
there is a {\it curvature singularity at} $x=\infty$; according
to \cite{HE}, or the classification scheme introduced in
\cite{ES}, it is a ``p.p. curvature singularity'', or a
``$C^0$ curvature singularity'', respectively. Geodesic
observers moving along the time-like geodesics (\ref{E5.33e}),
(\ref{E5.34}) inevitably  reach the singularity $x=+\infty$
in a finite  proper time ($\Delta\tau\sim\int_{x_0}^\infty
x^{-7/2}\, dx \sim [x^{-5/2}]_{x_0}^\infty=const<\infty$),
whereas observers $x=x_0<0$ moving along (\ref{E5.32}) escape
(there are no timelike observers $x=x_0>0$).

 On the other hand,  the region $x=0$ can not be reached
by any time-like observer (cf. (\ref{E5.31b}) where
$\dot x^2\sim -x^2/\beta^2$ as $x\rightarrow0$ would be a
contradiction). This indicates that $x=0$ represents the null
and/or spacelike infinity. Indeed, the transformation
\begin{eqnarray}
  \eta&=&{\beta\cos(T/\beta)}/{\cal D}\ , \ \quad
  x   \ =\ {\beta\cos\chi}/{\cal D}     \ , \label{E6.16} \\
  y   &=&{\beta\sin\chi\cos\vartheta}/{\cal D}  \ ,\quad
  z   \ =\ {\beta\sin\chi\sin\vartheta\cos\varphi}/{\cal D} \ , \nonumber
\end{eqnarray}
where $\eta=(u-v)/\sqrt2$, $z=(u+v)/\sqrt2$ and
${\cal D}=\sin(T/\beta)+\sin\chi\sin\vartheta\sin\varphi$,
brings the metric (\ref{E4.0})
into the form
\begin{eqnarray}
ds^2&=&\frac{\beta^2}{\cos^2 \chi}
  \left\{-\frac{dT^2}{\beta^2}+d\chi^2+\sin^2\chi(d\vartheta^2+
       \sin^2\vartheta\, d\varphi^2)\right\} \label{E6.17} \\
&+&\frac{\beta^5}{2}\frac{\cos\chi}
  {{\cal D}^5}\Bigg\{
  -[1+\cos(T/\beta-\varphi)\sin\chi\sin\vartheta]\frac{dT}{\beta}
  +\sin(T/\beta-\varphi)\cos\chi\sin\vartheta\, d\chi \nonumber\\
&&\qquad +\sin(T/\beta-\varphi)\sin\chi\cos\vartheta\, d\vartheta
-\sin\chi\sin\vartheta[\cos(T/\beta-\varphi)+\sin\chi\sin\vartheta]
  \,d\varphi \Bigg \}^2\ . \nonumber
\end{eqnarray}
This form of the Kaigorodov space-time shows explicitly that
the metric approaches asymptotically  the anti-de Sitter metric in
standard global coordinates as $x\rightarrow0$, i.e., $\chi\rightarrow\pm\pi/2$
(cf. \S 5.2 in \cite{HE} where $\cosh r=1/\cos\chi$).
In the literature, these coordinates are used for the construction of the
Penrose diagram of the anti-de Sitter space-time ---
choosing a conformal factor $\Omega=\frac{1}{\beta}\cos\chi$,
the boundary of the anti-de Sitter manifold given by $\Omega=0$
($\chi=\pm\pi/2$) represents null and spacelike
infinity which  can be thought of as a {\it timelike} surface since
$\Omega_{,\alpha} \Omega^{,\alpha}>0$ at $\Omega=0$.
Using the same conformal factor for the Kaigorodov
space-time (\ref{E6.17}) we conclude that the boundary $\Omega=0$
corresponding to $x=0$ represents
an `anti-de Sitter-like' scri having the topology $R\times S^2$.
Therefore, the Kaigorodov space-time is {\it weakly asymptotically anti-de
Sitter} according to the definition given in \cite{AM}. This is
not true, of course, in regions where ${\cal D}=0$ representing
singularities $x=\pm\infty$.

The fact that $x=0$ is  an infinity for null and spacelike
observers is supported by the asymptotic behavior of the
geodesics described by the equation (\ref{E5.33}). For spacelike
geodesics ($\epsilon=+1$) we get  $|x|\sim \exp(\Delta\tau/\beta)$
as  $x\rightarrow0$. Similarly, null geodesics ($\epsilon=0$)
are asymptotically $x\sim (\Delta\tau)^{-2/5}$
(cf. (\ref{E5.33f})), or $|x|\sim (\Delta\tau)^{-1}$ (for
$2AC+B^2<0$). Therefore, the boundary $\Omega=0$ is reached only
at infinite value of the affine parameter $\tau$ of null and
spacelike geodesics.

An important consequence of this is that the Kaigorodov
space-time (\ref{E4.0}) splits, in fact, into two disjointed
regions $x>0$ and $x<0$ (corresponding to different signs in
(\ref{E4.1})) which can not mutually communicate. These two regions are
quite different. For example, the singularity $x=+\infty$ is
reached by most of the timelike, null, and spacelike geodesics in
$x>0$, the singularity $x=-\infty$ is reached by {\it only
one} (spacelike) geodesic $x(\tau)=-\exp{[(\tau-\tau_0)/\beta]}$
(cf. (\ref{E5.33a})) as $\tau\rightarrow\infty$.
Also, the geodesics (\ref{E5.32})
given by $x=x_0$ are spacelike or null in $x>0$ whereas they are
timelike and null in $x<0$. Moreover, $g_{\mu\nu}\xi^\mu_{(3)}
\xi^\nu_{(3)}=\beta^2 x$ so that the Killing vector
${\bf \xi}_{(3)}=\partial_u$ (see (\ref{EE10})) is spacelike for $x>0$
but timelike for $x<0$. This implies that the Kaigorodov
space-time is {\it stationary} in $x<0$ with $u$ being a time coordinate.

Note that, surprisingly, the stationarity does not exclude
the presence of gravitational radiation. For example, standard
``homogeneous'' (or ``plane'') {\it pp}-waves \cite{KSMH, EK}
--- a `textbook model' of exact gravitational waves ---
are given by the metric
$ ds^2= 2 d\xi d\bar\xi-2dudv-(g+\bar g) du^2$,
where $g=A(u)\xi^2$. Considering the simplest case, $A=1$, and
introducing real spacelike coordinates $x$ and $y$ by
$\sqrt2\xi=x+iy$, we get
$ds^2= dx^2+dy^2-2dudv-(x^2-y^2) du^2$. This space-time
 is also stationary in the regions where
$|x|>|y|$ since the Killing vector ${\bf \xi}=\partial_u$ is
timelike: it represents a radiation with a constant amplitude
rather than a periodic-like gravitational wave.

\section {Concluding remarks}

Our study of the Siklos class of exact solutions indicates that
a reasonable physical interpretation of these space-times can be given
if one investigates the equation of geodesic deviation in the suitable
frame. As in the linearized theory, exact  waves of the
Siklos type manifest themselves by typical effects on particle
motions (transversality and specific polarization properties).
The  space-times describe exact gravitational waves propagating
in the anti-de Sitter universe. Somewhat surprising is that,
due to the presence of a negative cosmological constant,
the direction of propagation of the waves rotates.

The analysis provided independent arguments for the
interpretation suggested by Siklos \cite{Siklos}
of the space-times (\ref{E2.1}) as `Lobatchevski {\it pp}-waves'
(Siklos proved  that they admit a foliation of totally
geodesic two-dimensional spacelike surfaces of constant negative
curvature which are the wave surfaces of gravitational waves)
and also for Ivor Robinson's suggestion \cite{Ozs2}
that the Kaigorodov solution may be interpreted
as a `plane-fronted gravitational wave against the anti-de Sitter
background' (since the group of isometries is five parametric).
Therefore, the Siklos space-times --- and the Kaigorodov
solution in particular ---  can be understood as natural cosmological
($\Lambda<0$) analogues of the {\it pp}-waves in the flat
universe.

We hope that the studied space-times can be used in numerical relativity
as the test beds for numerical codes aimed to understand
realistic situations in more general cosmological contexts.

\section*{Acknowledgments}

I would like to thank Ji\v r\'\i\  Bi\v c\'ak for many very enlightening
discussions and Jerry Griffiths for reading the manuscript. I am grateful to
Daniel Finley, III for the warm hospitality in Albuquerque where this
work was started. I also acknowledge the support of grants
GACR-202/96/0206 and GAUK-230/96 of the Czech Republic and Charles University.

\end{document}